\documentclass[english,conference]{IEEEtran}
\usepackage[T1]{fontenc}
\usepackage[latin9]{inputenc}
\usepackage{amsmath}
\usepackage{amssymb}
\usepackage{graphicx}
\usepackage{esint}
\graphicspath{{../../figures/}}

\makeatletter

\usepackage{amsfonts}
\usepackage{url}
\usepackage{rotating}
\usepackage{color,soul}   
\date{}   

\makeatother

\usepackage{babel}
\begin{document}

\title{Statistical Characterization and Mitigation of NLOS Errors in UWB
Localization Systems}

\author{\IEEEauthorblockN{Francesco Montorsi} \IEEEauthorblockA{Department
of Information Engineering\\
 University of Modena and Reggio Emilia\\
 Modena, Italy 0039-0592056323\\
 Email: francesco.montorsi@unimore.it} \and \IEEEauthorblockN{Fabrizio
Pancaldi} \IEEEauthorblockA{Department of Science and \\
 Methods for Engineering\\
 University of Modena and Reggio Emilia\\
 Reggio Emilia, Italy 0039-0592056323\\
 Email: fabrizio.pancaldi@unimore.it}\and \IEEEauthorblockN{Giorgio
M. Vitetta} \IEEEauthorblockA{Department of Information Engineering\\
 University of Modena and Reggio Emilia\\
 Modena, Italy 0039-0592056323\\
 Email: giorgio.vitetta@unimore.it}}

\maketitle
\global\long\def\w{\operatorname{w}}
\global\long\def\u{\operatorname{u}}

\begin{abstract}
In this paper some new experimental results about the statistical
characterization of the \emph{non-line-of-sight} (NLOS) bias affecting
\emph{time-of-arrival} (TOA) estimation in \emph{ultrawideband} (UWB)
wireless localization systems are illustrated. Then, these results
are exploited to assess the performance of various \emph{maximum-likelihood}
(ML) based algorithms for joint TOA localization and NLOS bias mitigation.
Our numerical results evidence that the accuracy of all the considered
algorithms is appreciably influenced by the LOS/NLOS conditions of
the propagation environment.\end{abstract}
\begin{IEEEkeywords}
Ultra Wide Band (UWB), Radiolocalization, Bias mitigation.
\end{IEEEkeywords}

\section{Introduction\label{sec:Introduction}}

Wireless localization in harsh communication environments (e.g., in
a building where wireless nodes are separated by concrete walls and
other obstacles) can be appreciably affected by direct path attenuation
and NLOS conditions. Various solutions for NLOS error mitigation in
UWB environments are available in the technical literature \cite{wed_model},
\cite{nlos_marano}, \cite{decarli}, \cite{venkatesh_buehrer}. A
simple deterministic model, dubbed \emph{wall extra delay}, is proposed
in \cite{wed_model} to estimate the bias introduced by walls. A non-parametric
support vector machine is employed in \cite{nlos_marano} for joint
bias mitigation and channel status detection; this approach exploits
multiple features extracted from received signals in a non-statistical
fashion. A few classification algorithms for LOS/NLOS detection are
compared in \cite{decarli}, where it is shown that the best solution
is offered by a statistical strategy based on the joint \emph{probability
density function} (pdf) of the delay spread and the kurtosis extracted
from the received signals. Finally, in \cite{venkatesh_buehrer} statistical
models for the \emph{time of arrival} (TOA), the \emph{received signal
strength} (RSS) and the \emph{root mean square delay spread} (RDS)
are developed and an iterative estimator for bias mitigation is devised.

The contribution of this paper is twofold. In fact, first of all,
the problem of joint statistical modeling of multiple features extracted
from a database of waveforms acquired in a TOA-based localization
system is investigated. Note that, as far as we know, in the technical
literature only univariate models for bias mitigation have been proposed
until now (e.g., see \cite{wed_model}, \cite{venkatesh_buehrer}).
The use of multiple signal features in UWB localization systems has
been investigated in \cite{decarli} for channel state detection only
and in \cite{guvenc}, where, however, the considered features (namely,
the kurtosis, the mean excess delay and the delay spread) are modelled
as independent random variables. The second contribution is represented
by a performance comparison of various \emph{maximum-likelihood} (ML)
estimators for TOA-based localization. In particular, unlike other
papers (e.g., see \cite{nlos_marano}, \cite{decarli}) we illustrate
some numerical results referring to the accuracy of different localization
strategies, rather than to the bias removal on a single radio link.

The remaining part of this paper is organized as follows. In Section
\ref{sec:Experimental-activities} some information about our UWB
experimental campaign and about the features extracted from the acquired
data are provided. In Section \ref{sec:Localization-algorithms} some
estimation algorithms for UWB radiolocalization are described, whereas
their performance is compared in Section \ref{sec:Numerical-results}.
Finally, some conclusions are given in Section \ref{sec:Conclusions}.

\section{Experimental setup\label{sec:Experimental-activities}}

\subsection{Measurement arrangement}

A measurement campaign has been conducted by our research group in
2010; it is worth mentioning that various databases providing a collection
of sampled UWB waveforms acquired in experimental campaigns and useful
for assessing the performance of localization algorithms are already
available (e.g., see \cite{wprb_database}, \cite{usc_database}).
However, our database has been specifically generated to assess the
correlation between the NLOS bias error and various features extracted
from the received signals, as it will become clearer in the next Paragraph. 

\begin{figure}
\begin{centering}
\includegraphics[scale=0.7]{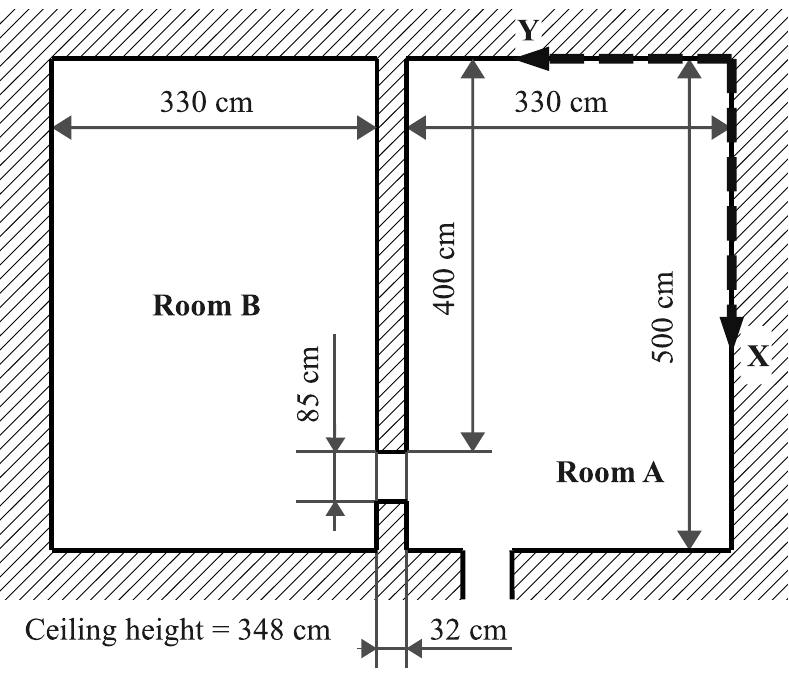}
\par\end{centering}

\caption{Map of the environment of our experimental campaign.\label{fig:environment}}
\end{figure}

All the measured data were acquired by means of two FCC-compliant
PulsON220 radios commercialised by TimeDomain and were collected in
a database. Such devices are equipped with omnidirectional antennas,
are characterized by a -10 dB bandwidth and a central frequency equal
to 3.2 GHz and 4.7 GHz, respectively, and perform two-way TOA ranging;
they also allow to store the digitised received waveforms (a sampling
frequency of 24.2 GHz and 14 bits per sample are used). Our measurement
campaign consisted of two phases. First, the transmitter was placed
in a given room (room A in Fig. \ref{fig:environment}) and the receiver
in an adjacent room (room B in Fig. \ref{fig:environment}) separated
from the room A by a wall having thickness $t_{wall}=32$ cm (NLOS
condition); in addition, the transmit antenna was kept fixed, whereas
the receive antenna was placed in $N_{acq}^{\text{\tiny{NLOS}}}=174$
distinct vertices of a dense square grid (the distance between a couple
of nearest vertices was equal to 21 cm). In the second phase of our
measurement campaign the transmitter and receiver were both placed
in room B (LOS condition); then, the transmit antenna was kept fixed,
whereas the receive antenna was moved on the same dense grid as in
the first phase, to acquire the UWB signal in $N_{acq}^{\text{\tiny{LOS}}}=105$
distinct vertices. The distance between the two antennas varied between
1 m and about 5 m; larger distances were not taken into consideration
since we were interested in indoor ranging only. It is important to
note that the choice of the measurement scenarios described above
is motivated by the fact that the UWB signals experienced similar
propagation in both phases. 

For each grid position, besides the acquired waveform, a TOA estimate
(evaluated by the PulsON220 devices) and the actual transmitter-receiver
distance (evaluated by means of a metric tape with an accuracy better
than 1 cm) were also stored in the database. The TOA estimates have
been used to provide a common time frame to all the acquired waveforms;
this has made possible the estimation of the mean excess delay and
of other signal statistics in the signal processing phase.

\subsection{Statistical modeling of signal features\label{sub:Statistical-results}}

The model 
\begin{equation}
\tau_{i}=\frac{d_{i}}{c_{0}}+b_{i}+w_{i}\label{eq:experimental_sigmodel}
\end{equation}
was adopted for the $i$-th link for both LOS and NLOS conditions.
Here, $\tau_{i}$ denotes the TOA estimated by the PulsON220 devices,
based on energy thresholds and a go-back technique (see \cite{toa_estimators}
for further details), $c_{0}$ is the speed of light, $d_{i}$ denotes
the \emph{distance} between the transmitter and the receiver, $b_{i}$
is the NLOS \emph{bias} (in seconds) affecting the TOA measurement
(the values taken on by this random parameter are always positive
for NLOS links and null for LOS links%
\footnote{In this case the \emph{probability density function} (pdf) of $b_{i}$
is $f_{b}(b)=\delta(b)$.%
}) and $w_{i}\sim\mathcal{N}(0,\sigma_{w,i}^{2})$ is the \emph{measurement
noise}; in addition, the expression $\sigma_{w,i}^{2}=\gamma\sigma_{n}^{2}d_{i}^{\beta}$
is adopted for the variance of the measurement noise, where $\gamma$
is a parameter depending on both the specific TOA estimator employed
in the ranging measurements and on various parameters of the physical
layer, and $\beta$ is the \emph{path-loss exponent} (a known and
fixed value is assumed for this parameter in both LOS and NLOS conditions
\cite{venkatesh_buehrer}).

In the following we focus on the problem of estimating the bias $b_{i}$
(affecting the TOA estimate $\tau_{i}$ (\ref{eq:experimental_sigmodel}))
from a set of $N_{f}=6$ different ``features\textquotedblright{}
$\left\{ x_{i,j}\text{, }j=0\text{, }1\text{, }...\text{, }5\right\} $
extracted from the received waveform $r_{i}(t)$. In particular, like
in \cite{nlos_marano}, the following features have been evaluated
for the set of the received waveforms:
\begin{enumerate}
\item the \emph{maximum signal amplitude} $x_{i,0}=r_{max,i}\triangleq\max_{t}|r_{i}(t)|$;
\item the \emph{mean excess delay} $x_{i,1}=\tau_{m,i}\triangleq\int_{0}^{\infty}t\frac{|r_{i}(t)|^{2}}{\varepsilon_{i}}dt$
(the parameter $\varepsilon_{i}$ is defined below);
\item the \emph{delay spread }$x_{i,2}=\tau_{ds,i}\triangleq\int_{0}^{\infty}(t-\tau_{m,i})^{2}\frac{|r_{i}(t)|^{2}}{\varepsilon_{i}}dt$;
\item the \emph{energy} $x_{i,3}=\varepsilon_{i}\triangleq\int_{0}^{\infty}|r_{i}(t)|^{2}dt$;
\item the \emph{rise time} $x_{i,4}=t_{rise,i}\triangleq\min$ $\left\{ t:|r_{i}(t)|/\max_{t}|r_{i}(t)|>0.9\right\} -\min\left\{ t:|r_{i}(t)|/\max_{t}|r_{i}(t)|>0.1\right\} $;
\item the \emph{kurtosis} $x_{i,5}=\kappa_{i}\triangleq\frac{1}{\sigma_{|r|}^{4}T}\int_{T}\left(|r_{i}(t)|-\mu_{|r|}\right)^{4}dt$,
where $\mu_{|r|}\triangleq\frac{1}{T}\int_{T}|r_{i}(t)|dt$, $\sigma_{|r|}^{2}\triangleq\frac{1}{T}\int_{T}(|r_{i}(t)|-\mu_{|r|})^{2}dt$
and $T$ denotes the \emph{observation time}. 
\end{enumerate}
In the following the set $\left\{ r_{i}(t)\right\} $ of received
waveforms is modelled as a random process, so that the above mentioned
features form a set of correlated random variables; in addition, all
of them are statistically correlated with the TOA bias. The last consideration
is confirmed by the numerical results of Table \ref{tab:corr_values_b_xj},
which lists the absolute values of the correlation coefficients of
the previously described features with the estimated TOA\ bias for
both the LOS and the NLOS scenarios. From these results it can be
easily inferred that not all the considered features are equally useful
to estimate the TOA bias. For this reason and to simplify our statistical
analysis, we restricted the set of features to $\{x_{i,0}$, $x_{i,1}$,
$x_{i,2}\}$ ($N_{f}=3$), which collects the parameters exhibiting
a strong correlation with the bias in the NLOS scenario.

\begin{table}
\begin{centering}
\begin{tabular}{|@{}c@{}|@{}c@{}|@{}c@{}|@{}c@{}|@{}c@{}|@{}c@{}|@{}c@{}|c|} \hline  \begin{tabular}{c} Correlation\tabularnewline with $b_{i}$ \tabularnewline \end{tabular}  & \multicolumn{1}{c|}{$x_{i,0}$} & \multicolumn{1}{c|}{$x_{i,1}$} & \multicolumn{1}{c|}{$x_{i,2}$} & \multicolumn{1}{c|}{$x_{i,3}$} & \multicolumn{1}{c|}{$x_{i,4}$} & \multicolumn{1}{c|}{$x_{i,5}$} & $d_{i}$\tabularnewline \hline \hline  NLOS case & 0.795 & 0.852 & 0.894 & 0.641 & 0.454 & 0.644 & 0.609\tabularnewline \hline  LOS case & 0.602 & 0.586 & 0.129 & 0.666 & 0.586 & 0.119 & 0.629\tabularnewline \hline \end{tabular}
\par\end{centering}

\caption{Absolute value of the correlation coefficient between $b_{i}$ and
each feature of the set $\{x_{i,j}$, $j=0$, $1$, $....$, $5\}$.\label{tab:corr_values_b_xj}}
\end{table}

Note that, in principle, the bias $b_{i}$ is not influenced by the
distance $d_{i}$, since it depends only on the thickness of the walls
(or of other obstacles) encountered by the transmitted signal during
its propagation; this is not true, however, for the above mentioned
triple of signal features (see {[}1{]} for further details). Generally
speaking, it is useful to derive a TOA bias estimator which is not
influenced by the transmitter-receiver distance $d_{i}$. Therefore,
in the attempt of removing the dependence of the features $\{x_{i,0}$,
$x_{i,1}$, $x_{i,2}\}$ on the link distance, we developed the models
$x_{i,0}=r_{max,i}=r_{max}^{0}-r_{max}^{m}d_{i}\text{,}$ $x_{i,1}=\tau_{m,i}=\tau_{m}^{m}d_{i}$
and $x_{i,2}=\tau_{ds,i}=\tau_{ds}^{0}+\tau_{ds}^{m}d_{i}$ on the
basis of our experimental results and, in particular, on the basis
of the estimates of the joint pdf's $\{f_{b,x_{0}}$, $f_{b,x_{1}}$,
$f_{b,x_{2}}\}$ referring to the three possible couples $(b_{i}$,
$x_{i,j})$ with $j=0$, $1$, $2$ and $i=0,1,...,N_{acq}-1$, where
$N_{acq}=N_{acq}^{\text{\tiny{NLOS}}}+N_{acq}^{\text{\tiny{LOS}}}$;
here $r_{max}^{0}$, $\tau_{m}^{m}$ and $\tau_{ds}^{m}$ are random
variables, whereas $\tau_{ds}^{0}$ and $r_{max}^{m}$ are deterministic
parameters having known values (all are assumed independent of $d_{i}$
and thus also from the link index $i$). Given these models, the vector
of distance-independent features $\tilde{\mathbf{x}}\triangleq\left[r_{max}^{0},\tau_{m}^{m},\tau_{ds}^{m}\right]^{T}$
can be evaluated from its distance-dependent counterpart%
\footnote{In the rest of the document, the subscript $i$ has been omitted for
simplicity when not strictly necessary.%
} $\mathbf{x}=\left[r_{max},\tau_{m},\tau_{ds}\right]^{T}$and can
be used in place of it. However it turns out that parameters of $\tilde{\mathbf{x}}$
are less correlated with $b_{i}$ than those of $\mathbf{x}$; for
this reason, we decided to take into consideration both for localization
purposes (see Sec. \ref{sub:parameterization_choice}).

\begin{figure}
\begin{centering}
\includegraphics[width=8cm]{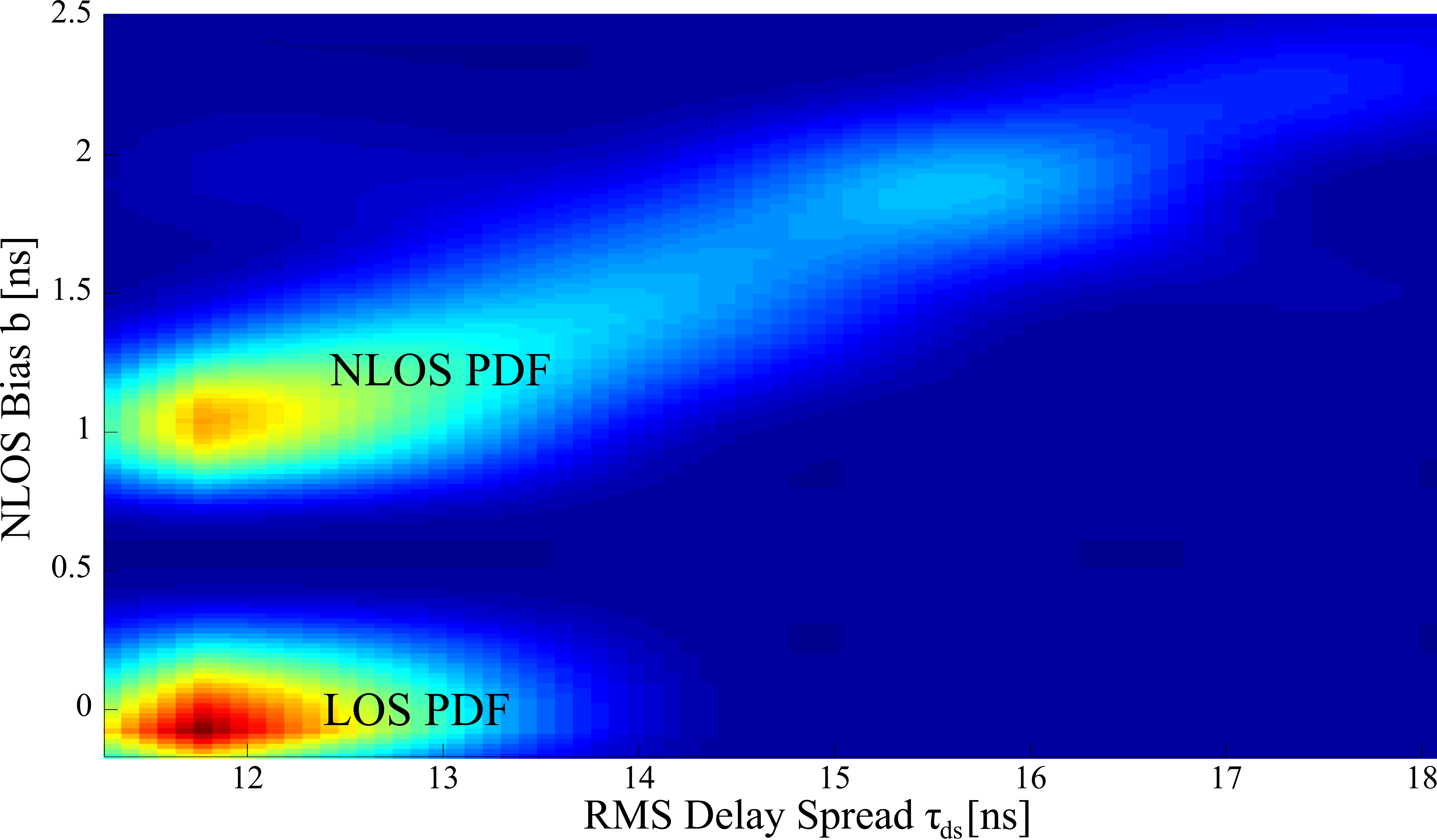}
\par\end{centering}

\caption{Estimated joint pdf of the estimated TOA bias and the delay spread
in NLOS conditions (above) and LOS conditions (below).\label{fig:pdf_2d_bias_taurms}}

\end{figure}

A complete statistical characterization of the estimated bias  and
of the $3$ related signal features we consider is provided by the
joint pdf $f_{b,r_{max},\tau_{m},\tau_{ds}}(\cdot)$ or, equivalently,
by the joint pdf $f_{b,r_{max}^{0},\tau_{m}^{m},\tau_{ds}^{m}}(\cdot)$
(both can be estimated from the acquired data). Some marginalizations,
like the one shown in Fig. \ref{fig:pdf_2d_bias_taurms}, reveal interesting
aspects: 1) a significant (limited) correlation between these parameters
is found in the NLOS (LOS) case; 2) the null region exhibited by the
estimated pdf's is due to the fact that the TOA bias cannot take on
values in the interval $[0,\: t_{wall}/c_{0}]$, where $t_{wall}$
is the thickness of the wall obstructing the direct path; 3) large
values of the TOA bias are unlikely since they are associated with
small incidence angles of the transmitted signal on the obstructing
wall.

\section{Localization algorithms \label{sec:Localization-algorithms}}

\subsection{Introduction}

In this Section we develop various algorithms for two-dimensional
localization in a UWB network composed by $N_{a}$ anchors with \emph{known}
positions $\mathbf{z}_{i}^{a}\triangleq\left[x_{i}^{a},\, y_{i}^{a}\right]^{T}\in\mathbb{R}^{2}$,
$i=0,...,N_{a}-1$ and by a single node (dubbed mobile station, MS,
in the following) with \emph{unknown} position $\boldsymbol{\theta}\triangleq[x,\, y]^{T}\in\mathbb{R}^{2}$.
Any couple of the given $(N_{a}+1)$ devices can operate in a LOS
(NLOS) condition with probability $P_{\text{\tiny{LOS}}}$ ($1-P_{\text{\tiny{LOS}}}$).
The localization algorithms described below try to mitigate the effects
of the NLOS bias error and aim at generating an estimate $\hat{\boldsymbol{\theta}}$
of $\boldsymbol{\theta}$ minimizing the \emph{mean square error}
(MSE) $\mathbb{E}_{\mathbf{\hat{\boldsymbol{\theta}}}}\left\{ ||\mathbf{\hat{\boldsymbol{\theta}}}-\mathbf{\boldsymbol{\theta}||}^{2}\right\} $.
It is also important to point out that localization algorithms developed
for NLOS scenarios usually consist of two steps. In fact, first the
NLOS bias is estimated for each involved link and is used to remove
the bias contribution in the acquired data; then the new data set
is processed by a \emph{least-square} (LS) procedure generating an
estimate of $\mathbf{\boldsymbol{\theta}}$ (e.g., see \cite{nlos_marano},
\cite{decarli}, \cite{venkatesh_buehrer}, \cite{guvenc}). In the
following instead, implicit estimation of the bias for each link is
adopted; this approach is motivated by the fact that the estimation
of the bias for the $i$-th link can benefit from the information
acquired from the other ($N_{a}-1$) links; note that in \cite{nlos_marano}
and \cite{guvenc} bias mitigation performed in a link-by-link fashion
is exploited to assign a weight in a \emph{weighted least-square}
(WLS) step but such an approach is heuristic and differs from our
ML-based approach.

\subsection{Maximum Likelihood Estimation}

If the links between the MS and the $N_{a}$ different anchors are
assumed mutually independent, the ML estimation strategy of the unknown
MS position $\mathbf{\boldsymbol{\theta}}$, given a TOA estimate
and a set of additional signal features for each link, can be formulated
as $\mathbf{\hat{\boldsymbol{\theta}}}=\arg\underset{\mathbf{\tilde{\boldsymbol{\theta}}}}{\max}\ln\prod_{i=0}^{N_{a}-1}f_{\mathbf{\tau},\mathbf{x}}(\tau_{i},\mathbf{x}_{i};\mathbf{\tilde{\boldsymbol{\theta}}})$,
where $\mathbf{\tilde{\boldsymbol{\theta}}}=[\tilde{x},\,\tilde{y}]^{T}\in\mathbb{R}^{2}$
denotes the MS trial position, $\tau_{i}$ and $\mathbf{x}_{i}$ are
the TOA and $N_{f}$-dimensional signal vector collecting the received
signal features acquired for the $i$-th link and $f_{\tau,\mathbf{x}}(\cdot,\mathbf{\cdot};\mathbf{\tilde{\boldsymbol{\theta}}})$
is the joint pdf of the TOA and the vector of features parameterized
by the trial position $\mathbf{\tilde{\boldsymbol{\theta}}}$. In
the following Paragraphs the impact of various possible options for
the ML strategy are investigated.

\subsubsection{ML estimation strategy for different sets of observed data\label{sub:dimensionality_choice}}

in this Paragraph two different options are considered for the set
of features processed by the ML strategy.

\paragraph{Option A}

in this case it is assumed that the vector of features referring to
the $i$-th link is $\mathbf{x}_{i}=[r_{max,i},\tau_{m,i},\tau_{ds,i}]^{T}$;
this choice is motivated by the large correlation between these random
variables and the link bias $b$ (see Paragraph \ref{sub:Statistical-results}).
Then, the joint PDF $f_{\mathbf{\tau},\mathbf{x}}(\tau_{i},\mathbf{x}_{i};\mathbf{\tilde{\boldsymbol{\theta}}})$
appearing in the ML strategy can be expressed as (see Eq. (\ref{eq:experimental_sigmodel})):
\begin{equation}
f_{\tau,\mathbf{x}}(\tau_{i},\mathbf{x}_{i};\tilde{\boldsymbol{\theta}})=(f_{b,\tilde{\mathbf{x}}}\otimes f_{w})\left(\tau_{i}-\frac{d_{i}(\mathbf{\tilde{\boldsymbol{\theta}}})}{c_{0}},\tilde{\mathbf{x}}_{i}\right),\label{eq:4d_ml_pdf}
\end{equation}
where 
\begin{equation}
\tilde{\mathbf{x}}_{i}=\left[r_{max,i}+r_{max}^{m}d_{i}(\mathbf{\tilde{\boldsymbol{\theta}}}),\frac{\tau_{m,i}}{d_{i}(\mathbf{\tilde{\boldsymbol{\theta}}})},\frac{\tau_{ds,i}-\tau_{ds}^{0}}{d_{i}(\mathbf{\tilde{\boldsymbol{\theta}}})}\right]^{T}
\end{equation}
(because of the models previously described for $r_{max}^{0}$, $\tau_{m}^{m}$
and $\tau_{ds}^{m}$), $d_{i}(\mathbf{\tilde{\boldsymbol{\theta}}})$
is the distance between the $i$-th anchor and the MS trial position
and $f_{b,\tilde{\mathbf{x}}}\otimes f_{w}$ denotes the convolution
between the joint pdf $f_{b,\tilde{\mathbf{x}}}(\cdot)$ and the observation
noise pdf $f_{w}(w)=(2\pi\sigma_{w}^{2})^{-1/2}\exp(-w^{2}/(2\sigma_{w}^{2}))$.

Note that the shape of the function $f_{b,\tilde{\mathbf{x}}}(\cdot)$
under the \emph{hypothesis of} LOS \emph{conditions} ($H_{\text{\tiny{LOS}}}$
event) is substantially different from that found in NLOS conditions
($H_{\text{\tiny{NLOS}}}$ event) \cite{venkatesh_buehrer}; for this
reason, we estimated this function in both cases applying the procedure
described in \cite{wed_model} to the data collected in our measurement
campaign; this led to two distinct multidimensional histograms, which
approximate the pdf's $f_{b,\tilde{\mathbf{x}}}\left(b,\tilde{\mathbf{x}}_{i}|H_{\text{\tiny{LOS}}}\right)$
and $f_{b,\tilde{\mathbf{x}}}\left(b,\tilde{\mathbf{x}}_{i}|H_{\text{\tiny{NLOS}}}\right)$
with a certain accuracy depending on: a) the quantity of acquired
data; b) the sizes $\Delta b$, $\Delta r_{max}$, $\Delta\tau_{m}$
and $\Delta\tau_{ds}$ of the quantization bins adopted in the generation
of the histograms. Note that these sizes need to be accurately selected,
since large bins imply a coarse approximation of pdf's, whereas excessively
small bins require a huge amount of data.

Given an estimate of the above mentioned couple of pdf's, the required
pdf $f_{b,\tilde{\mathbf{x}}}(\cdot)$ can be trivially evaluated
using the law of total probability and the $P_{\text{\tiny{LOS}}}\triangleq\Pr\{H_{\text{\tiny{LOS}}}\}$
and $P_{\text{\tiny{NLOS}}}\triangleq\Pr\{H_{\text{\tiny{NLOS}}}\}=1-\Pr\{H_{\text{\tiny{LOS}}}\}$
probabilities when they are available, or assuming $P_{\text{\tiny{LOS}}}=P_{\text{\tiny{NLOS}}}=0.5$
if no \emph{a priori} information about the LOS/NLOS conditions are
available.

\paragraph{Option B}

in this case the set of features employed in ML estimation consists
of a single element, namely the delay spread (which exhibits the largest
correlation with the NLOS bias; see Table \ref{tab:corr_values_b_xj}),
so that $\mathbf{x}_{i}=\tau_{ds,i}$ and the pdf $f_{\mathbf{\tau},\mathbf{x}}(\tau_{i},\mathbf{x}_{i};\mathbf{\tilde{\boldsymbol{\theta}}})$
of the ML strategy becomes (see (\ref{eq:4d_ml_pdf}))
\begin{equation}
f_{\tau,\mathbf{x}}(\tau_{i},\mathbf{x}_{i};\tilde{\boldsymbol{\theta}})=(f_{b,\tau_{ds}^{m}}\otimes f_{w})\left(\tau_{i}-\frac{d_{i}(\mathbf{\tilde{\boldsymbol{\theta}}})}{c_{0}},\frac{\tau_{ds,i}-\tau_{ds}^{0}}{d_{i}(\mathbf{\tilde{\boldsymbol{\theta}}})}\right)\label{eq:2d_ml_pdf}
\end{equation}
Like in the previous case, the pdf $f_{b,\tau_{ds}^{m}}(\cdot)$ has
to be estimated in the LOS and NLOS scenarios (see Fig. \ref{fig:pdf_2d_bias_taurms})
from the data acquired in our measurement campaign. Note that this
option leads to a ML localization algorithm which is substantially
simpler than that proposed in the analysis of option A.

\subsubsection{Parameterization of the observations\label{sub:parameterization_choice}}

the ML strategies developed above are based on the joint pdf's $f_{b,r_{max}^{0},\tau_{m}^{m},\tau_{ds}^{m}}(\cdot)$
and $f_{b,\tau_{ds}^{m}}(\cdot)$ which refer to a set of \emph{distance-independent
parameters}. Since the parameters $r_{max}^{0}$, $\tau_{m}^{m}$
and $\tau_{ds}^{m}$ exhibit a lower correlation with $b$ than their
distance-dependent counterparts $r_{max}$, $\tau_{m}$, $\tau_{ds}$,
we believe that the use of the joint PDF
\begin{equation}
f_{\tau,\mathbf{x}}(\tau_{i},\mathbf{x}_{i};\mathbf{\tilde{\boldsymbol{\theta}}})=(f_{b,\mathbf{x}}\otimes f_{w})\left(\tau_{i}-\frac{d_{i}(\tilde{\boldsymbol{\theta}})}{c_{0}},\mathbf{x}_{i}\right)
\end{equation}
in place of (\ref{eq:4d_ml_pdf}) deserves to be investigated (similar
comments hold for (\ref{eq:2d_ml_pdf})).

\subsubsection{Estimation of joint pdf's\label{sub:postprocessing_choice}}

as already explained above, the joint pdf's involved in the proposed
ML localization strategies can be easily estimated from the acquired
data using a simple procedure based on dividing the space of observed
data in a set of bins of proper size. Such a procedure generates an
histogram, which, unluckily, entails poor localization performance
if employed as it is, because of the relatively small number of bins
(adopted to avoid empty bins). To mitigate this problem either interpolation
followed by low-pass filtering can be applied to raw experimental
histograms (this leads to \emph{interpolated-histogram} \emph{estimators})
or the raw data can be fitted with analytical functions (this results
in \emph{fitted-histogram} \emph{estimators}).

\subsection{Iterative Estimation\label{sub:Iterative-Estimator}}

\begin{figure}
\begin{centering}
\includegraphics[scale=0.6]{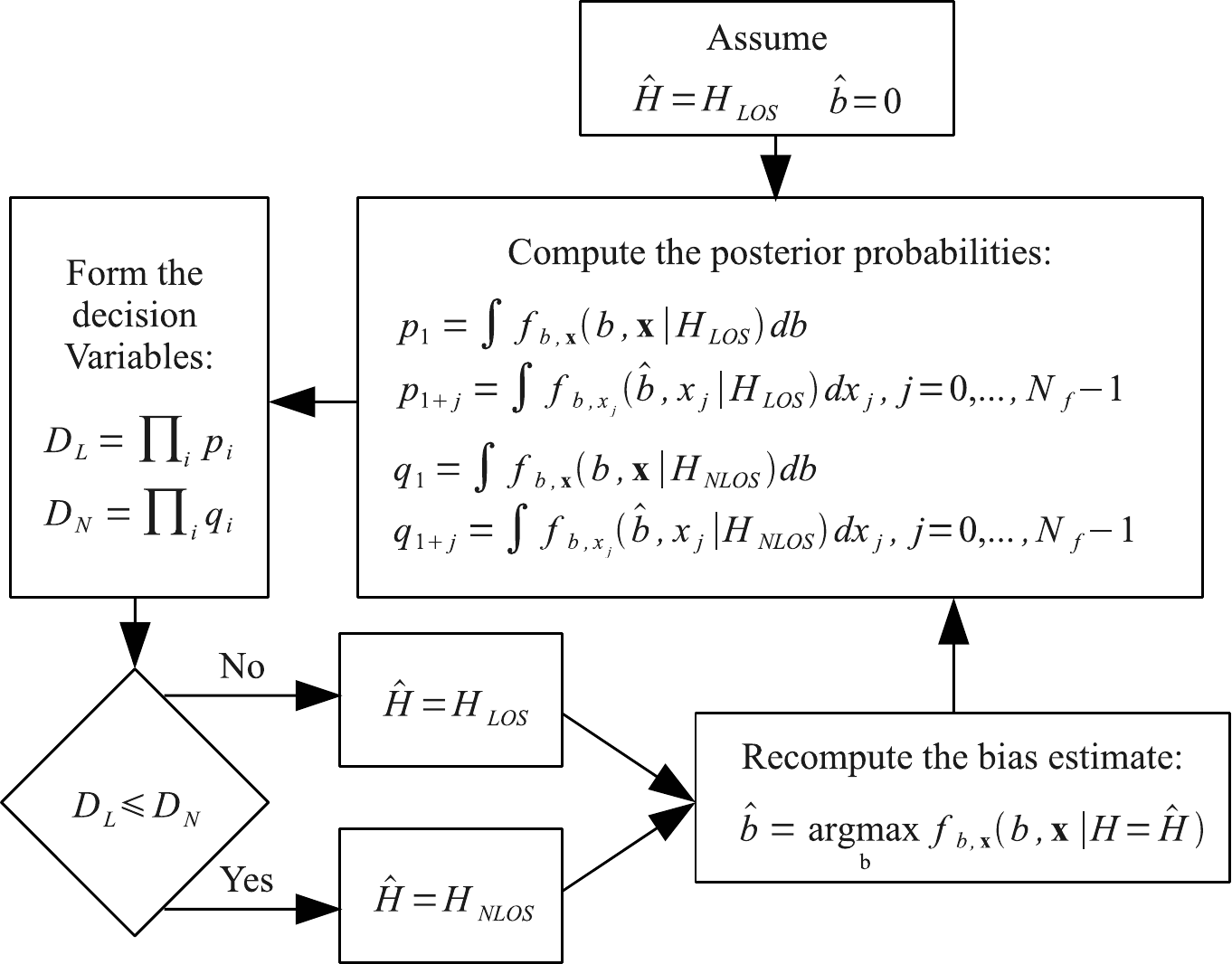}
\par\end{centering}

\caption{Flow diagram of the proposed iterative estimator.\label{fig:iterative_block_diagram}}

\end{figure}

Recently, an iterative estimator of both the channel state (i.e.,
LOS or NLOS conditions) and the NLOS bias $b$ has been proposed in
\cite{venkatesh_buehrer}. In Fig. \ref{fig:iterative_block_diagram}
the flow diagram of a modified version of this iterative algorithm,
employing the joint pdf's described above and extracted from our experimental
database, is proposed. Note that this algorithm operates in a link-by-link
fashion (in the diagram of Fig. \ref{fig:iterative_block_diagram}
the link index $i$ has been omitted to ease the reading).

\section{Numerical results\label{sec:Numerical-results}}

Our experimental database has also been exploited to assess the RMSE
performance of the proposed algorithms for localization and NLOS bias
mitigation schemes via computer simulations. In our simulations the
MS coordinates are always $(0;0)$; then, following \cite{nlos_marano},
for the $i$-th anchor a received waveform from either the LOS database
(with probability $P_{\text{\tiny{LOS}}}$) or from the NLOS database
(with probability $P_{\text{\tiny{NLOS}}}$) was drawn randomly and
was associated with the position $\mathbf{z}_{i}^{a}=\left(d_{i}\sin(2\pi\frac{i-1}{N_{a}});d_{i}\cos(2\pi\frac{i-1}{N_{a}})\right)$,
where $d_{i}$ is the distance measured for the selected waveform.
Note that these waveforms already include the experimental noise and
thus no simulated noise was imposed on the waveforms (so that the
signal-to-noise ratio is the experimental one). Finally, the parameter
$N_{a}$ has been set to $3$ (worst case which still theoretically
allows unambiguous localization).

\begin{figure}
\begin{centering}
\includegraphics[scale=0.65]{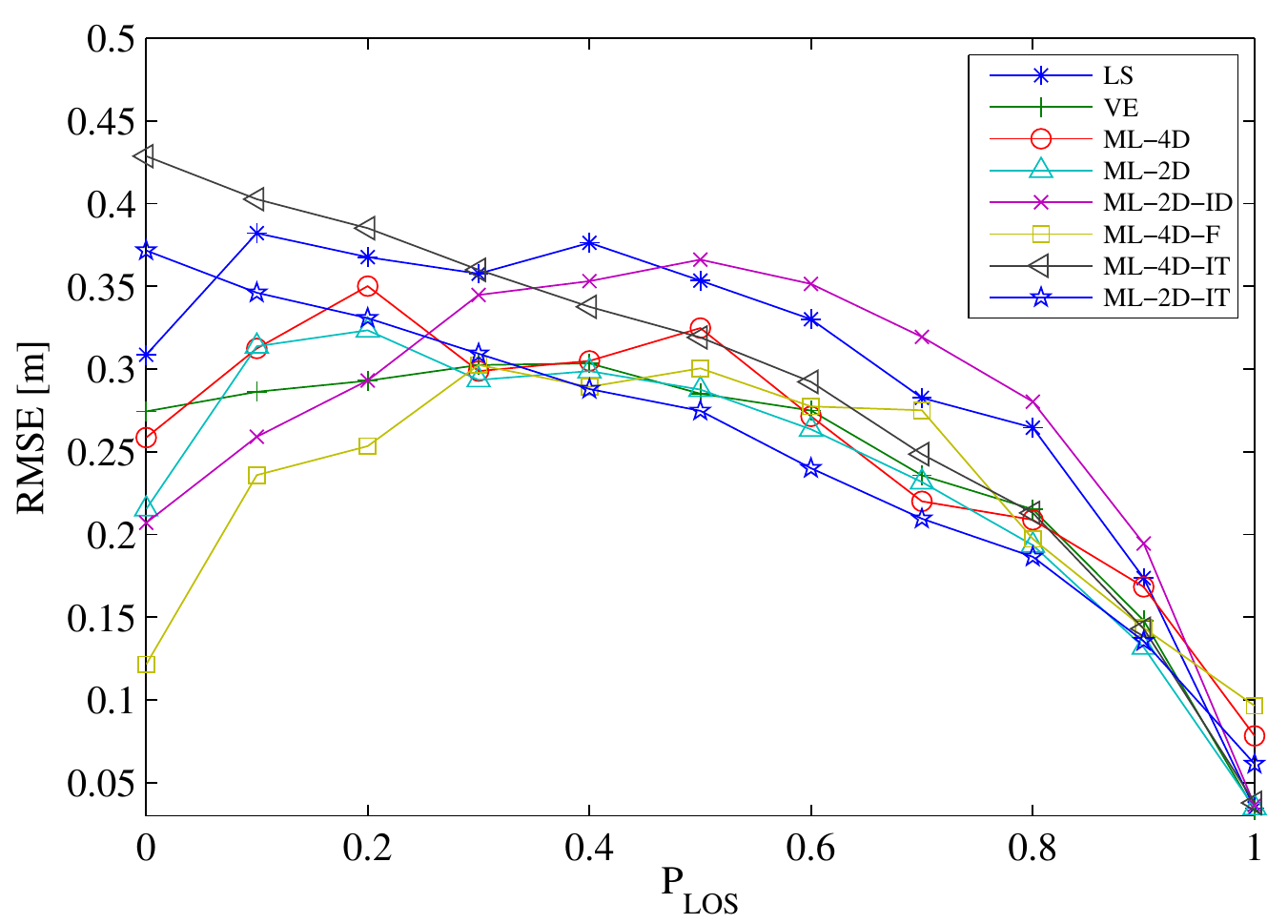}
\par\end{centering}

\caption{RMSE performance versus $P_{\text{\tiny{LOS}}}$ offered by various
localization algorithms.\label{fig:num_results}}
\end{figure}

The RMSE performance of the following algorithms has been evaluated:
\begin{enumerate}
\item \textbf{LS} - A standard LS estimator for LOS environments; the estimation
strategy can be expressed as $\hat{\mathbf{\boldsymbol{\theta}}}=\arg\min_{\mathbf{\tilde{\boldsymbol{\theta}}}}\sum_{i=0}^{N_{a}-1}(c_{0}\tau_{i}-d_{i}(\mathbf{\tilde{\boldsymbol{\theta}}}))^{2}$.
\item \textbf{VE} -\emph{ }A LS estimator exploiting TOA measurements corrected
by the algorithm proposed by Venkatesh and Buehrer in \cite{venkatesh_buehrer};
this algorithm relies on a statistical modeling of the propagation
environment based on our experimental database.
\item \textbf{ML-4D} - A ML estimator based on (\ref{eq:4d_ml_pdf}) and
employing an interpolated histogram, in the likelihood function, referring
to a distance-dependent parameterization.
\item \textbf{ML-2D} - A ML estimator based on (\ref{eq:2d_ml_pdf}) and
employing an interpolated histogram, in the likelihood function, referring
to a distance-dependent parameterization.
\item \textbf{ML-2D-ID} - A ML estimator based on (\ref{eq:2d_ml_pdf})
and employing an interpolated, histogram in the likelihood function,
referring to a distance-\emph{independent} parameterization.
\item \textbf{ML-4D-F} - A ML estimator based on (\ref{eq:4d_ml_pdf}) and
employing a \emph{fitted} histogram, in the likelihood function, referring
to a distance-dependent parameterization.
\item \textbf{ML-4D-IT - }A\textbf{ }LS estimator based on (\ref{eq:4d_ml_pdf})
and exploiting TOA measurements corrected by the modified \emph{iterative}
algorithm illustrated in Section \ref{sub:Iterative-Estimator}.
\item \textbf{ML-2D-IT - }A\textbf{ }LS estimator based on (\ref{eq:2d_ml_pdf})
and exploiting TOA measurements corrected by the modified \emph{iterative}
algorithm illustrated in Section \ref{sub:Iterative-Estimator}. 
\end{enumerate}
In estimating the RMSE performance of the ML algorithms listed above
the likelihood functions were always evaluated at the vertices of
a square grid characterized by a step size equal to $10$ mm. Some
numerical results are compared in Fig. \ref{fig:num_results}, which
illustrates the RMSE performance versus the probability $P_{\text{\tiny{LOS}}}$.
These results evidence that:
\begin{enumerate}
\item The simple LS algorithm is outperformed by all the other algorithms
when $P_{\text{\tiny{LOS}}}\leq0.9$; this is due to the fact that
this strategy does not try to mitigate NLOS bias.
\item The VE algorithm performs well at the cost of a reasonable complexity,
but offers limited bias mitigation when $P_{\text{\tiny{LOS}}}=0$;
in this case the ML-2D, ML-2D-ID and ML-4D-F estimators perform much
better.
\item The exploitation of a large set of received signal features does not
necessarily allow to achieve better accuracy than a subset of them
(see the curves referring to ML-2D and ML-4D estimators); this is
due to the fact that the correlation between the different couples
of extracted features is typically large, so that they provide strongly
correlated information about the NLOS bias.
\item Distance-dependent parameterization provides better accuracy (see
the curves referring to the ML-2D and ML-2D-ID estimators); this can
be related to the fact that $\tau_{ds}^{m}$ is less correlated with
the NLOS bias than its distance-dependent counterpart $\tau_{ds}$.
\item The ML-4D-F estimator performs better than the ML-4D estimator in
NLOS conditions; this means that the use of fitted histograms entails
an improvement of localization accuracy.
\end{enumerate}

\section{Conclusions\label{sec:Conclusions}}

In this paper various UWB localization techniques processing multiple
features extracted from the received signal to mitigate the problem
of NLOS bias have been described and their accuracy has been assessed
exploiting the experimental data acquired in an measurement campaign.
Our results evidence that: a) a restricted set of features has to
be employed; b) the use of distance-dependent features and of fitted
histograms provides better performance than that offered by distance-independent
features and interpolated histograms.

\section*{Acknowledgements}

The authors would like to thank Prof. Marco Chiani and Prof. Davide
Dardari (both from the University of Bologna, Italy) for lending us
the UWB devices employed in our measurement campaign and the PhD students
Alessandro Barbieri and Fabio Gianaroli for their invaluable help
in our experimental work. Finally, the authors wish to acknowledge
the activity of the Network of Excellence in Wireless COMmunications
(NEWCOM++, contract n. 216715), supported by the European Commission
which motivated this work.


\begin{thebibliography}{1}

\bibitem{wed_model}
D.~Dardari, A.~Conti, J.~Lien, and M.~Z. Win, ``The effect of cooperation on
  localization systems using {UWB} experimental data,'' {\em EURASIP J. Adv.
  Signal Process}, vol.~2008, Jan. 2008.

\bibitem{nlos_marano}
S.~Maran\'{o}, W.~Gifford, H.~Wymeersch, and M.~Z. Win, ``{NLOS} identification
  and mitigation for localization based on {UWB} experimental data,'' {\em IEEE
  J. Sel. Areas Commun.}, vol.~28, pp.~1026 --1035, Sept. 2010.

\bibitem{decarli}
N.~Decarli, D.~Dardari, S.~Gezici, and A.~A. D'Amico, ``{LOS/NLOS} detection
  for {UWB} signals: A comparative study using experimental data,'' in {\em
  Proc. of the 5th IEEE International Symposium on Wireless Pervasive Computing
  (ISWPC 2010)}, pp.~169 --173, May 2010.

\bibitem{venkatesh_buehrer}
S.~Venkatesh and R.~Buehrer, ``Non-line-of-sight identification in
  ultra-wideband systems based on received signal statistics,'' {\em IET
  Microwaves, Antennas Propagation}, vol.~1, pp.~1120 --1130, Dec. 2007.

\bibitem{guvenc}
I.~G\"{u}ven\c{c}, C.-C. Chong, F.~Watanabe, and H.~Inamura, ``{NLOS}
  identification and weighted least-squares localization for {UWB} systems
  using multipath channel statistics,'' {\em EURASIP J. Adv. Signal Process},
  vol.~2008, Jan. 2008.

\bibitem{wprb_database}
``{WPR.B} database.'' Available online at \url{http://www.vicewicom.eu}.

\bibitem{usc_database}
J.-Y. Lee, ``{USC} ranging test database.'' Available online at
  \url{http://ultra.usc.edu/uwb_database/ranging_test.htm}.

\bibitem{toa_estimators}
D.~Dardari, C.-C. Chong, and M.~Z. Win, ``Threshold-based time-of-arrival
  estimators in {UWB} dense multipath channels,'' {\em IEEE Trans. Commun.},
  vol.~56, pp.~1366 --1378, Aug. 2008.

\end{thebibliography}
\end{document}